\documentclass[a4paper,11pt]{article}
\pdfoutput=1 
\usepackage{slashed}
\usepackage{color}
\usepackage{rotating}
\usepackage{cancel}
\usepackage{geometry}
\usepackage{hyperref}
\usepackage{graphicx}
\usepackage{dcolumn}
\usepackage{bm}
\usepackage{amsmath}
\usepackage[numbers, square, comma, sort&compress]{natbib}

\newcommand{\gcusp}{\gamma_{\rm cusp}}

\newcommand{\taun}{\mathcal{T}_N}
\newcommand{\tauzero}{\mathcal{T}_{0}}

\newcommand{\nn}{\nonumber}
\newcommand{\half}{\frac{1}{2}}
\newcommand{\eps}{\epsilon}
\newcommand{\mO}{\mathcal{O}}

\begin{document}
\begin{titlepage}
\begin{flushright}
LA-UR-18-24011\\
TUM-HEP-1137/18\\
1804.06358\\ 
\today 
\\  
\end{flushright}

\vskip 25mm

\begin{center}
\Large\bf{ Next-to-Next-to-Leading Order   $N$-Jettiness Soft Function \\
for $tW$ Production}
\end{center}

\vskip 8mm

\begin{center}

\textsc{ Hai Tao Li$^{a}$, Jian Wang$^b$}\\
\vspace{10mm}
\textit{$^a$Los Alamos National Laboratory, Theoretical Division, Los Alamos, NM 87545, USA } \\ 
\vspace{1mm}
\textit{$^b$Physik Department T31,  Technische Universit\"at M\"unchen, James-Franck-Stra\ss e~1,
D--85748 Garching, Germany} \\ 

\end{center}

\vspace{10mm}

\abstract{
We calculate the $N$-jettiness soft function  for $tW$ production  up to next-to-next-to-leading order in QCD,
which is an important ingredient of the $N$-jettiness subtraction method for predicting 
the differential cross sections of massive colored particle productions. 
The divergent parts of the results have been checked using the renormalization group equations controlled by the soft anomalous dimension. 
}

\end{titlepage}

\section{Introduction}

Precise calculation of cross sections for the processes at the Large Hadron Collider~(LHC) or future high-energy hadron colliders
is crucial for testing the Standard Model (SM) and for searching for new physics.
In the last a few years, there is a burst of fully differential next-to-next-to-leading order (NNLO) results 
for a large number of processes in the SM; see a recent review in ref.~\cite{Heinrich:2017una}.  
One of the main difficulties in the higher-order QCD calculations is to develop a systematical method to deal with the infrared singularities
caused by double real emissions.
The $N$-jettiness subtraction \cite{Boughezal:2015dva, Gaunt:2015pea} has proven to be successful in computing the NNLO differential cross sections of
processes with jets, for example, $W/Z/H/\gamma+j$ \cite{Boughezal:2015dva,Boughezal:2015aha,Boughezal:2015ded,Campbell:2016lzl}.
This subtraction method is based on the 
soft-collinear effective theory (SCET)~\cite{Bauer:2000ew,Bauer:2000yr,Bauer:2001ct,Bauer:2001yt,Beneke:2002ph},
which is an effective theory of QCD in the infrared regions.
The $N$-jettiness $\mathcal{T}_N$ is an observable, proposed in \cite{Stewart:2010tn},
to describe the event shape of jet processes or processes with initial-state hadrons,
a generalization of thrust at lepton colliders and beam thrust at hadron colliders \cite{Stewart:2009yx}.
The application of this observable to the NNLO calculations has been explored extensively
for massive quark decay \cite{Gao:2012ja},
and differential cross sections of processes at both  hadron colliders \cite{Boughezal:2015dva,Gaunt:2015pea,Boughezal:2015aha,Boughezal:2015ded,Berger:2016oht,Campbell:2016lzl,Heinrich:2017bvg}
and electron-hadron colliders~\cite{Berger:2016inr,Abelof:2016pby}.
It is also used as a jet resolution variable in combining higher-order resummation with  NLO calculations and parton showers \cite{Alioli:2012fc}.
However, for more complicated processes, e.g., involving one massive and two massless partons,
the results  are still missing.

The $N$-jettiness event shape variable is defined by~\cite{Stewart:2010tn}
\begin{align}
\label{eq:tau}
\taun = \sum_k \min_i\left\{n_i \cdot q_k \right\}~,
\end{align}
where $n_i$ $(i=a,b,1,...,N)$ are light-like reference vectors representing the moving directions of massless external particles,
and $q_k$ denotes the momentum of soft or collinear partons.
Note that eq.(\ref{eq:tau}) seems different from the original definition in ref.\cite{Stewart:2010tn}
because the variable $\taun$ in our definition is of mass dimension one while
that in ref.\cite{Stewart:2010tn} is dimensionless.
But they are actually the same up to a constant factor $Q$ after replacing 
$n_i$ by $2q_i/Q$.
In the infrared divergent regions, the observable  $\mathcal{T}_N \to 0$,
and  the cross section is approximated by \cite{Stewart:2009yx,Stewart:2010tn}
\begin{align}
\label{eq:sigma}
    \frac{d\sigma}{d\taun} \propto  \int H\otimes B_1 \otimes B_2 \otimes S \otimes \bigg(\prod_{n=1}^{N}J_n \bigg)~.
\end{align}
Here the hard function $H$ encodes all the  information about hard scattering.
The beam functions $B_i, ~(i=1,2),$ describe the perturbative and non-perturbative contributions  from initial state, 
and have been obtained up to NNLO~\cite{Stewart:2010qs,Berger:2010xi,Gaunt:2014xga,Gaunt:2014cfa}.
The jet function $J_n$ describes the final-state jet with a fixed invariant mass and has been calculated at NNLO~\cite{Becher:2006qw,Becher:2010pd}. The soft function $S$ contains soft interactions between all colored particles.
It has been studied up to NNLO for massless processes~\cite{Jouttenus:2011wh,Kelley:2011ng,Monni:2011gb, Boughezal:2015eha,Kang:2015moa,Campbell:2017hsw}.

The differential cross section for any observable $\mO$ is given by
\begin{align}
\label{eq:sigmatot}
    \frac{d\sigma}{d\mO}= \frac{d\sigma}{d\mO}\Big|_{\taun<\Delta} + \frac{d\sigma}{d\mO}\Big|_{\taun>\Delta} ~,
\end{align}
where a small cut-off parameter $\Delta$ on the right-hand side is imposed.
For the NNLO calculations the first term on the right-hand side at the leading power can be obtained
by expanding  eq.~(\ref{eq:sigma}) to the second order of the strong coupling $\alpha_s$.
The second term, due to the phase-space constraint, can be dealt with the standard NLO subtraction method 
for the process with an extra parton in the final state.

The extension of the $N$-jettiness subtraction to more complicated processes
requires the calculation of the corresponding soft and hard functions.
We have calculated the $N$-jettiness soft function for one massive colored particle production up to NNLO in ref.~\cite{Li:2016tvb},
where we assume that it is produced at rest.
In this paper, we present the result for more general situations, i.e., the massive colored particle can carry any possible momentum. 
Our result can be used to construct the $N$-jettiness subtraction terms for $tW$ production at hadron colliders. 

This paper is organized as follows.
In section~\ref{sec:factor}, we briefly introduce the definition of the soft function in terms of soft Wilson lines.
In section~\ref{sec:Renor}, we study the renormalization group (RG) equation of the soft function
and thus derive the structure of the soft function.
We provide the details of the techniques in our calculations in section \ref{sec:tech}.
Then, in section~\ref{sec:results}, we present the numerical results of the NLO
and NNLO soft functions and compare the divergent terms with the predictions from RG equation.
We conclude in section~\ref{sec:conc}.

\section{\label{sec:factor}Definition of the soft function}

In this section we first discuss the kinematics and the factorization of the cross section for $tW$ production. 
Then we present the definition of soft function. 

We consider the process
\begin{align}
    P_1 + P_2 \to t/\bar{t} + W^\pm + X~,
\end{align}
where $P_1$ and $P_2$ denote incoming hadrons, $t/\bar{t}$ and $W^\pm$  represent the top/anti-top quark  and the $W$-boson in the final state, respectively.  And $X$ includes any unobserved final state.
The partonic process at leading order (LO) for $tW^-$ production is
\begin{align}
\label{eq:process}
    b(p_1) + g(p_2) \to t(p_3)+W^-(p_4)~.
\end{align}
It is  convenient to introduce two light-like vectors
\begin{align}
\label{eq:nnb}
     n^\mu&=(1,0,0,1), \qquad \bar{n}^\mu=(1,0,0,-1)~.
\end{align}
Any momentum can be decomposed as $p^{\mu}=(p^+,p^-,p_{\perp})$
with $p^+= p\cdot n, p^- =p \cdot \bar{n}$.
The momenta given in eq.~(\ref{eq:process}) 
can be written in the partonic center-of-mass frame as
\begin{align}
     p_1^{\mu} = \frac{\sqrt{\hat{s}}}{2}n^{\mu}~,\quad  p_2^{\mu} = \frac{\sqrt{\hat{s}}}{2} \bar{n}^{\mu}, \quad
     p_3^\mu =m_t v^\mu~,
\end{align}
where $v^2=1$. Specifically, we parameterize $v$ by two variables, i.e., $\beta_t$ and $\theta_t$,
which measure the magnitude and the direction of the velocity,
\begin{align}
v^+ = \frac{1-\beta_t \cos \theta_t}{\sqrt{1-\beta_t^2}},\quad
v^- = \frac{1+\beta_t \cos \theta_t}{\sqrt{1-\beta_t^2}},\quad
|v_{\perp}|= \frac{\beta_t \sin \theta_t}{\sqrt{1-\beta_t^2}},
\end{align}
where $\beta_t=\sqrt{1-m_t^2/E_t^2}$ with $E_t$ the top quark energy.
The 0-jettiness event shape variable in this process is defined as
\begin{align} \label{eq:tau-def}
     \mathcal{\tau} \equiv \tauzero = & \sum_{k} \min\{n\cdot q_k, \bar{n}\cdot q_k \}~.
\end{align}
Since $n\cdot q_k\equiv 2p_1\cdot q_k/\sqrt{\hat{s}}$ and $\bar{n}\cdot q_k\equiv 2p_2\cdot q_k/\sqrt{\hat{s}}$,
this definition of $\mathcal{\tau}$ is Lorentz invariant.
The explicit choice in eq.(\ref{eq:nnb}) just makes our calculation easier, 
but the final result is general and independent of this choice.

In the limit $\tau \ll \sqrt{\hat{s}}$, the final state contains no hard radiations, only soft and collinear radiations allowed. In this limit the cross section admits a factorised form, which can be derived in the framework of SCET. 
For $tW$ production, the collinear singularities, which are only associated with the initial partons, and the soft singularities are all properly regularised by $\tau$ defined in eq.~(\ref{eq:tau-def}).  
Compared with processes without massive colored particles, the only difference is the soft function and the hard function.
 Following \cite{Stewart:2009yx,Stewart:2010tn}, we write
\begin{multline}
\label{eq:fact}
    \frac{d\sigma}{dY d\tau} = \int d\Phi_2\frac{d\hat{\sigma}_0}{d\Phi_2}  \int dt_a dt_b  d\tau_s H(\beta_t,\cos\theta_t, \mu) B_1(t_a, x_a, \mu) B_2(t_b, x_b, \mu)
    \\
    \times S(\tau_s, \beta_t,\cos\theta_t,\mu)\delta\left(\tau-\tau_s - \frac{t_a+t_b}{\sqrt{\hat{s}}}\right)\left(1+\mathcal{O}\left(\frac{\tau}{\sqrt{\hat{s}}}\right)\right)~,
\end{multline}
where $\int d\Phi_2$ is the  two-body phase space integral, $d\hat{\sigma}_0$ is the LO partonic differential cross section, $Y$ is the rapidity of the partonic colliding system in the laboratory frame,
the momentum fractions $x_a=\sqrt{\hat{s}/s} e^{Y}$ and $x_b=\sqrt{\hat{s}/s}e^{-Y}$ with $\sqrt{s}$ the collider energy,
and  $\mu$ is the renormalization scale.
In momentum space the soft function is defined as the vacuum matrix element
\begin{multline}\label{eq:soft}
    S(\tau,\beta_t,\cos\theta_t, \mu)=\sum_{X_s}
    \Big\langle 0 \Big|\mathbf{\bar{T}} Y_n^{\dagger}Y_{\bar{n}}Y_{v} \Big| X_s \Big\rangle
     \delta\bigg(\tau-\sum_{k} \min\left(n\cdot \hat{P}_k, \bar{n}\cdot \hat{P}_k\right)\bigg)
     \Big\langle X_s\Big| \mathbf{T} Y_nY_{\bar{n}}^{\dagger}Y_{v}^{\dagger}  \Big| 0 \Big\rangle,
\end{multline}
where $\mathbf{T}(\mathbf{\bar{T}})$ is the (anti-)time-ordering operator. And $Y_n$,  $Y_{\bar{n}}$ and $Y_{v}$ are the soft Wilson lines defined explicitly as \cite{Bauer:2001yt,Chay:2004zn,Korchemsky:1991zp}
\begin{align}
 Y_n(x) & = \mathbf{P} \exp\left( ig_s\int^0_{-\infty}ds\, n\cdot A^a_s(x+sn)\mathbf{T}^a\right) ,\\
 Y^{\dagger}_{\bar{n}}(x) & = \mathbf{\bar{P}} \exp\left( -ig_s\int^0_{-\infty}ds\, \bar{n}\cdot A^a_s(x+s \bar{n})\mathbf{T}^a\right), \\
 Y^{\dagger}_v(x) & = \mathbf{P} \exp\left( ig_s\int_0^{\infty}ds\, v\cdot A^a_s(x+sv)\mathbf{T}^a\right)
\end{align}
where  $\mathbf{P}$ and $\mathbf{\bar{P}}$ are the path-ordering and the anti-path-ordering operators.  $\hat{P}_k$ in eq.(\ref{eq:soft}) is the operator extracting the momentum of each soft emission. The purpose of this paper is to calculate the soft function defined above for $tW$ production up to NNLO accuracy.

\section{Renormalization} \label{sec:Renor}

In SCET the bare soft function in eq.(\ref{eq:soft}) contains ultra-violet divergences in perturbative calculations, which are cancelled by the counterterm defined in the standard renormalization procedure. The renormalized soft function is finite and can be used in the calculation of the cross section in eq.~(\ref{eq:fact}). The renormalization introduces  the  scale $\mu$ dependence in the soft function, as well as in the hard and beam function. Because of the fact that the physical cross section does not dependent on the intermediate scale, the RG equation of the soft function can be derived from the RG equations of the hard and beam function, which will be used to extract the anomalous dimension of the soft function.  Given the anomalous dimension the divergences in the bare soft function, as well as the scale dependence of the renormalized soft function, can be predicted. In this section we briefly discuss the renormalization of the soft function and the expression of the soft anomalous dimension.  
We work in $d=4-2\epsilon$ dimensional space-time.

Based on dimensional analysis, the bare soft function, in perturbation theory, can be written  as
\begin{align}\label{eq:baresoft}
     S(\tau,\beta_t,\cos\theta_t, \mu) = \delta(\tau) + \frac{1}{\tau} \sum_{n=1}^{\infty} \left(\frac{Z_{\alpha_s} \alpha_s}{4\pi}\right)^n \left(\frac{\tau}{\mu}\right)^{-2n\epsilon} s^{(n)}(\beta_t,\cos\theta_t)~,
\end{align}
where  we use renormalized strong coupling $\alpha_s$ and its renormalization factor $Z_{\alpha_s}=1-\beta_0 \alpha_s/(4\pi \epsilon)+\mathcal{O}(\alpha_s^2)$.
The soft function after the Laplace transformation can be written as 
\begin{align}\label{eq:lapsoft}
    \tilde{S}(L,\beta_t,\cos\theta_t,\mu) &= \int_{0}^{\infty} d\tau \exp\left( -\frac{\tau}{e^{\gamma_E}\mu e^{L/2}} \right) S(\tau,\beta_t,\cos\theta_t,\mu)
          \nonumber \\
        &= 1+\sum_{n=1}^{\infty} \left(\frac{Z_{\alpha_s}\alpha_s}{4\pi} \right)^n e^{-n(L+2\gamma_E)\epsilon}\Gamma(-2n\epsilon) s^{(n)}(\beta_t,\cos\theta_t)~.
\end{align}
Then the corresponding renormalized soft function $\tilde{s} $ is defined as
\begin{align}\label{eq:soft_ren}
     \tilde{s}(L,\beta_t,\cos\theta_t,\mu) = Z_s^{-1}(L,\beta_t,\cos\theta_t,\mu) \tilde{S}(L,\beta_t,\cos\theta_t,\mu)~,
\end{align}
where the renormalization factor $Z_s$ satisfies the differential equation
\begin{align}
    \frac{ d\ln Z_s(L,\beta_t,\cos\theta_t,\mu)} {d\ln\mu} = -\gamma_s(L,\beta_t,\cos\theta_t,\mu)
\end{align}
with $\gamma_s$  the anomalous dimension of the soft function. We will suppress the arguments of the renormalization factor, anomalous dimension and the soft function in the following text for convenience. 

Given the soft anomalous dimension  $\gamma_s$,
 following refs.~\cite{Becher:2009cu,Becher:2009qa}, the closed expression for $Z_s$ is derived and can be written as 
\begin{align}
     \ln Z_s &= \frac{\alpha_s}{4\pi}\left(\frac{\gamma_s^{(0)\prime}}{4\epsilon^2}+ \frac{\gamma_s^{(0)}}{2\epsilon} \right)
     + \left(\frac{\alpha_s}{4\pi}\right)^2 \left(-\frac{3 \beta_0 \gamma_s^{(0)\prime} }{16\epsilon^3} + \frac{\gamma_s^{(1)\prime} -4 \beta_0 \gamma_s^{(0)}}{16\epsilon^2}+ \frac{\gamma_s^{(1)}}{4\epsilon} \right)
     +\mathcal{O}(\alpha_s^3).
\end{align}
The expansion series and derivative of the soft anomalous dimension are given by
\begin{align}
     \gamma_s = \sum_{i=0} \left( \frac{\alpha_s}{4\pi}\right)^{i+1} \gamma_s^{(i)}
     \qquad \text{and}  \qquad \gamma_s^{(i)\prime} = \frac{d \gamma_s^{(i)}}{d\ln\mu}~.
\end{align}
From eq.~(\ref{eq:soft_ren}), we obtain the renormalized NLO and NNLO soft functions in Laplace space 
\begin{align}
\label{eq:s_div}
    \tilde{s}^{(1)} =& \tilde{S}^{(1)}-Z_s^{(1)} ~,
    \nonumber \\
    \tilde{s}^{(2)} =& \tilde{S}^{(2)}-Z_s^{(2)} - \tilde{S}^{(1)} Z_s^{(1)}+ Z_s^{(1)2}  -\frac{\beta_0}{\epsilon} \tilde{S}^{(1)}~.
\end{align}
Since the renormalized soft function is finite,  the divergent terms in the 
bare soft function $\tilde{S}$ is related to the renormalization factor $Z_s$ and can be derived from the above equations. 

As discussed before, the soft anomalous dimension $\gamma_s$ can be derived from the independence of the cross section on the renormalization scale $\mu$, 
\begin{align} \label{eq:s_rg}
  \frac{d\ln \tilde{s}} {d\ln \mu} =\gamma_s  = -\frac{d\ln H}{d\ln \mu} - \frac{d\ln \tilde{B}_1}{d\ln \mu} - \frac{d\ln\tilde{B}_2}{d\ln \mu}~,
\end{align}
where $ \tilde{B}_i$ is the beam function in Laplace space, of which the NLO and NNLO results can be found in  refs.~\cite{Stewart:2010qs,Berger:2010xi,Gaunt:2014xga,Gaunt:2014cfa}.
And  the  RG equation of the beam function is exactly the same as the evolution equation of the jet function to all orders~\cite{Stewart:2010qs},
\begin{align}\label{eq:beam}
     \frac{d\tilde{B}_i}{d\ln\mu} = \left(- \mathbf{T}_i\cdot \mathbf{T}_i \gcusp \Big( \ln \frac{s_{12}}{\mu^2} +L\Big)+ \gamma_B^i\right) \tilde{B}_i~,
\end{align}
where $\mathbf{T}_i$ is the  color generator associated with the $i$-th parton~\cite{Catani:1996jh,Catani:1996vz} and the anomalous dimension $\gamma_B^i$ can be found in refs.~\cite{Gaunt:2014xga,Gaunt:2014cfa}.

The RG equation for the hard function can be obtained from  refs.~\cite{Ferroglia:2009ep,Ferroglia:2009ii} where the two-loop divergences have been calculated for massive scattering amplitudes in non-abelian gauge theories. 
It is straightforward to organize the RG equation for the  hard function as
\begin{align}\label{eq:hard}
     \frac{d\ln H}{d\ln\mu} =& (\mathbf{T}_1\cdot \mathbf{T}_1+\mathbf{T}_2\cdot \mathbf{T}_2) \ln \frac{s_{12}}{\mu^2}
     -  \mathbf{T}_1\cdot \mathbf{T}_3 \gcusp \ln \frac{s_{13}^2}{s_{12} m_t^2}
     -  \mathbf{T}_2\cdot \mathbf{T}_3 \gcusp \ln \frac{s_{23}^2}{s_{12} m_t^2}
     \nn\\
&     +2\gamma^1 + 2\gamma^2 + 2\gamma^Q~
\end{align}
with $s_{12}=2p_1\cdot p_2 + i0,s_{13}=-2p_1\cdot p_3 + i0,s_{23}=-2p_2\cdot p_3 + i0$. 
The anomalous dimensions $\gamma^{1,2}$ and $\gamma^Q$,
associated with the initial- and final-state particles, can be found in refs. \cite{Ferroglia:2009ep,Ferroglia:2009ii} and references therein.

Inserting eqs.~(\ref{eq:beam}-\ref{eq:hard}) to eq.(\ref{eq:s_rg}), the anomalous dimension of the soft function is obtained 
\begin{align}\label{eq:gammas}
     \gamma_s =& (\mathbf{T}_1\cdot \mathbf{T}_1 + \mathbf{T}_2\cdot \mathbf{T}_2)\gcusp L
     + \mathbf{T}_1\cdot \mathbf{T}_3 \gcusp \ln \frac{s_{13}^2}{s_{12} m_t^2} 
     + \mathbf{T}_2\cdot \mathbf{T}_3 \gcusp \ln \frac{s_{23}^2}{s_{12} m_t^2}\nn\\&
     - 2 \gamma^Q
     - \gamma_B^1-\gamma_B^2 - 2\gamma^1-2\gamma^2~,
\end{align}
of which each ingredient is available up to NNLO.

\section{Techniques in calculation}\label{sec:tech}
In the calculation of the NLO and NNLO soft function,
we have to deal with one and two soft radiations, respectively.
The phase space integration is 
\begin{align}
\int \frac{d^d q}{(2\pi)^d}\delta^+(q^2)&=\frac{1}{(2\pi)^d}\frac{\Omega_{d-3}}{4}\int dq^+ dq^- (q^+ q^-)^{-\epsilon}\int_0^{\pi} d\phi_q \sin^{-2\eps}\phi_q \\
\int \frac{d^dq_1d^dq_2}{(2\pi)^{2d}} \delta^+(q_1^2)\delta^+(q_2^2)  &=\frac{1}{(2\pi)^{2d}}\frac{\Omega_{d-3}^2}{16} 
\int dq_1^+ dq^-_1 (q^+_1 q^-_1)^{-\epsilon} 
 \int dq^+_2 dq^-_2 (q^+_2 q^-_2)^{-\epsilon} \nn\\ & \quad
d\phi_{1} \sin^{-2\eps}\phi_{1}  d\phi_{2} \sin^{-2\eps}\phi_{2} \label{eq:phase2a} 
\end{align}
with 
\begin{align}
\Omega_{d-3}  &=  \frac{2\pi^{\half-\eps}}{\Gamma(\half-\eps)}  =  
2-3 \zeta(2) \epsilon ^2-\frac{14 \zeta (3) \epsilon ^3}{3}-\frac{15}{8} \zeta(4) \epsilon ^4+O\left(\epsilon ^5\right) .
\end{align}
The $\phi$ angle is measured in the frame with the top quark $\phi_t=0$.
More explicitly,
we choose 
\begin{align}
p_{3\perp}&=|p_{3\perp}|(0; 0, 1) ,  \nn\\
q_{1\perp}&=|q_{1\perp}|(0; \sin\phi_1, \cos\phi_1) ,  \nn\\
q_{2\perp}&=|q_{2\perp}|( \sin\phi_2 \sin\beta\hat{n}_{\eps};   \sin\phi_2 \cos\beta, \cos\phi_2 ).
\end{align}

For the integrand involving $1/(q_1\cdot q_2)$, the phase space integration is parameterize as 
\begin{align}
\int \frac{d^dq_1d^dq_2}{(2\pi)^{2d}} \delta^+(q_1^2)\delta^+(q_2^2)  &=\frac{1}{(2\pi)^{2d}}\frac{\Omega_{d-3}\Omega_{d-4}}{16} 
\int dq_1^+ dq^-_1 (q^+_1 q^-_1)^{-\epsilon} 
 \int dq^+_2 dq^-_2 (q^+_2 q^-_2)^{-\epsilon} \nn\\ & \quad
d\phi_{1} \sin^{-2\eps}\phi_{1}  d\phi_{12} \sin^{-2\eps}\phi_{12}  d\beta_{12} \sin^{-1-2\eps}\beta_{12}
\label{eq:phase2b}
\end{align}
with 
\begin{align}
\Omega_{d-4}  &=  \frac{2\pi^{-\eps}}{\Gamma(-\eps)}  =  \frac{-2\eps\pi^{-\eps} }{\Gamma(1-\eps)}.
\end{align}

At NLO, the measurement function is defined as
\begin{align}
     F(n,\bar{n},q) & =
     \delta(q^+-\tau)\Theta(q^--q^+)
      + \delta(q^--\tau)\Theta(q^+-q^-)~,
\end{align}
where  $q^+=q\cdot n$ and  $q^-=q\cdot \bar{n}$.
At NNLO, the measurement function is defined as
\begin{align}\label{eq:measure}
     F(n,\bar{n},q_1,q_2) &= \delta(q_1^++q_2^+-\tau)\Theta(q^-_1-q_1^+)\Theta(q_2^--q_2^+)
     \nonumber \\ &
      +\delta(q_1^++q_2^--\tau)\Theta(q^-_1-q_1^+)\Theta(q_2^+-q_2^-)
           \nonumber \\ &
      +\delta(q_1^-+q_2^--\tau)\Theta(q^+_1-q_1^-)\Theta(q_2^+-q_2^-)
           \nonumber \\ &
             +\delta(q_1^-+q_2^+-\tau)\Theta(q^+_1-q_1^-)\Theta(q_2^--q_2^+).
\end{align}
One can see that at NNLO the whole phase space is partitioned to four pieces.
We  label them as Region-I, Region-II, Region-III and Region-IV, respectively.

In the hemisphere with $q_i^+=\tau_i$, we parameterize $q_i^- =\tau_i/t_i$ with $t_i \in (0,1)$ and 
\begin{align}
dq_i^- = dt_i \frac{\tau_i}{t_i^2},\quad (q_i^+ q_i ^-)^{-\eps} = (\tau_i^2/t_i)^{-\eps}  .
\end{align}
And then all those singularities at NLO will appear as $\tau_i^{-1-2\eps}$ and $t_i^{-1+\eps}$.
In the end, we define 
\begin{align}
\tau_1 = \tau v, \quad \tau_2 =\tau\bar{v},
\end{align}
with $\bar{v}\equiv 1-v$. We have 
\begin{align}
\int d\tau_1 d\tau_2 \delta(\tau-\tau_1-\tau_2) =\tau \int_0^1 dv.
\end{align}
If the integrands do not involve $1/(q_1\cdot q_2)$, we perform the phase space integration straightforward after the parameterization.
For the  integrands involving $1/(q_1\cdot q_2)$, we have
\begin{align}
q_1\cdot q_2 
&= \half\frac{ \tau_1 \tau_2} {t_1 t_2 }\left[ t_2  + t_1 -2 \sqrt{t_1 t_2} \cos \phi_{12} \right] ,
\end{align}
in the Region-I or Region-III,
and
\begin{align}
q_1\cdot q_2
&= \half\frac{ \tau_1 \tau_2} {t_1 t_2 }\left[ 1  + t_1 t_2 -2 \sqrt{t_1 t_2} \cos \phi_{12} \right]  ,
\end{align}
in the Region-II or  Region-IV.
Here $\phi_{12}$ is the angle between $q_{1\perp}$ and $q_{2\perp}$.

In the Region-I and Region-III, the double-real corrections contain a new kind of singularities that appear when $t_1 =t_2$ and $\phi_{12}=0$. 
Following the method in ref.~\cite{Campbell:2017hsw}, we change the integration variables from ${\phi_2, \beta}$ to ${\phi_{12},\beta_{12}}$. 
All dependence on $\phi_2$ (such as $q_{2\perp}\cdot p_{3\perp}$) can be expressed in terms of $\phi_{12} $ and $\beta_{12}$,
\begin{align}
\cos\phi_2 = \cos\phi_1 \cos \phi_{12} - \sin \phi_1 \sin \phi_{12} \cos \beta_{12}.
\end{align}
The $\beta_{12}$ angle integration can be transformed by defining $\cos\beta_{12}=1-2x$,
\begin{align}
\int_0^{\pi} d\beta_{12} (\sin\beta_{12})^{-1-2\eps}&=2^{-1-2\eps} \int_0^1 dx [x(1-x)]^{-1-\eps}.
\label{eq:betaint}
\end{align}
Define $\cos\phi_{12} = 1-2z$, 
\begin{align}
q_1\cdot q_2 &= \half\frac{ \tau_1 \tau_2} {t_1 t_2 }\left[ (\sqrt{t_2}  -\sqrt{ t_1})^2 +4z\sqrt{t_1 t_2} \right]  \nn\\
&= \half\frac{ \tau_1 \tau_2} {t_1 t_2 }\frac{(t_2-t_1)^2}{\left[ (\sqrt{t_2}  -\sqrt{ t_1})^2 +4r \sqrt{t_1 t_2} \right]}
 .
\label{eq:q1q2}
\end{align}
By writing in this form, we have picked out the singular part as $t_2 \to t_1$.
The parameter $r$ is solved to be
\begin{align}
r =\frac{(\sqrt{t_2}-\sqrt{t_1})^2 (1 -z ) }{(\sqrt{t_2} -\sqrt{t_1})^2 + 4 z \sqrt{t_1 t_2}},
\end{align}
and the Jacobian is 
\begin{align}
\frac{dz}{dr} = -\frac{(t_2-t_1)^2}{\left[ (\sqrt{t_2}  -\sqrt{ t_1})^2 +4r \sqrt{t_1 t_2} \right]^2}.
\end{align}
The $\phi_{12}$ angular integration is given by
\begin{align}
\int_0^{\pi} d\phi_{12} \sin^{-2\eps}\phi_{12} &=4^{-\eps} \int_0^1 dz [z(1-z)]^{-\half-\eps}
\label{eq:z}\\
&=4^{-\eps} \int_0^1 dr [r(1-r)]^{-\half-\eps}
\frac{|t_2 - t_1|^{1-2\eps}}{\left[ (\sqrt{t_2}  -\sqrt{ t_1})^2 +4r \sqrt{t_1 t_2} \right]^{1-2\eps}}.
\end{align}
Combined with eq.(\ref{eq:q1q2}), we see that the singular part of $(q_1\cdot q_2)^{-1}\sim  |t_2 - t_1|^{-1-2\eps}$ 
and $(q_1\cdot q_2)^{-2}\sim  |t_2 - t_1|^{-3-2\eps}$.
However, we find that the coefficient of $(q_1\cdot q_2)^{-2}$ is proportional to $(t_1-t_2)^2$.
Now we divide the integration region of $t_1 ,t_2$ to two sectors, i.e.,
\begin{align}
1. \quad t_1> t_2&:  \quad t_2=t_1 (1-w), \quad w\in (0,1) ,  \nn\\
2. \quad t_1 < t_2&:  \quad t_1=t_2 (1-w), \quad w\in (0,1) .
\end{align}
In each sector, $|t_2 - t_1|$ has a definite sign and thus is easy to deal with.

In the Region-II and Region-IV, one can carry out the same procedure as above except 
the relation between $r$ and $z$ changes to 
\begin{align}
r =\frac{(1-\sqrt{t_1 t_2})^2 (1 -z ) }{(1 -\sqrt{t_1 t_2})^2 + 4 z \sqrt{t_1 t_2}}.
\end{align}
With this parametisation above, all the divergences can be extracted through the expansion,
\begin{align}
    x^{-1+n\eps} = \frac{1}{n\eps}\delta(x) + \left(\frac{1}{x}\right)_+ + n\eps \left(\frac{\ln x}{x}\right)_+ 
    + \cdots.
\end{align}
Then all the phase space integration can be performed numerically.

\section{Results of the soft function}\label{sec:results}
\subsection{NLO soft function}

\begin{figure}
    \centering
    \includegraphics[scale=0.35]{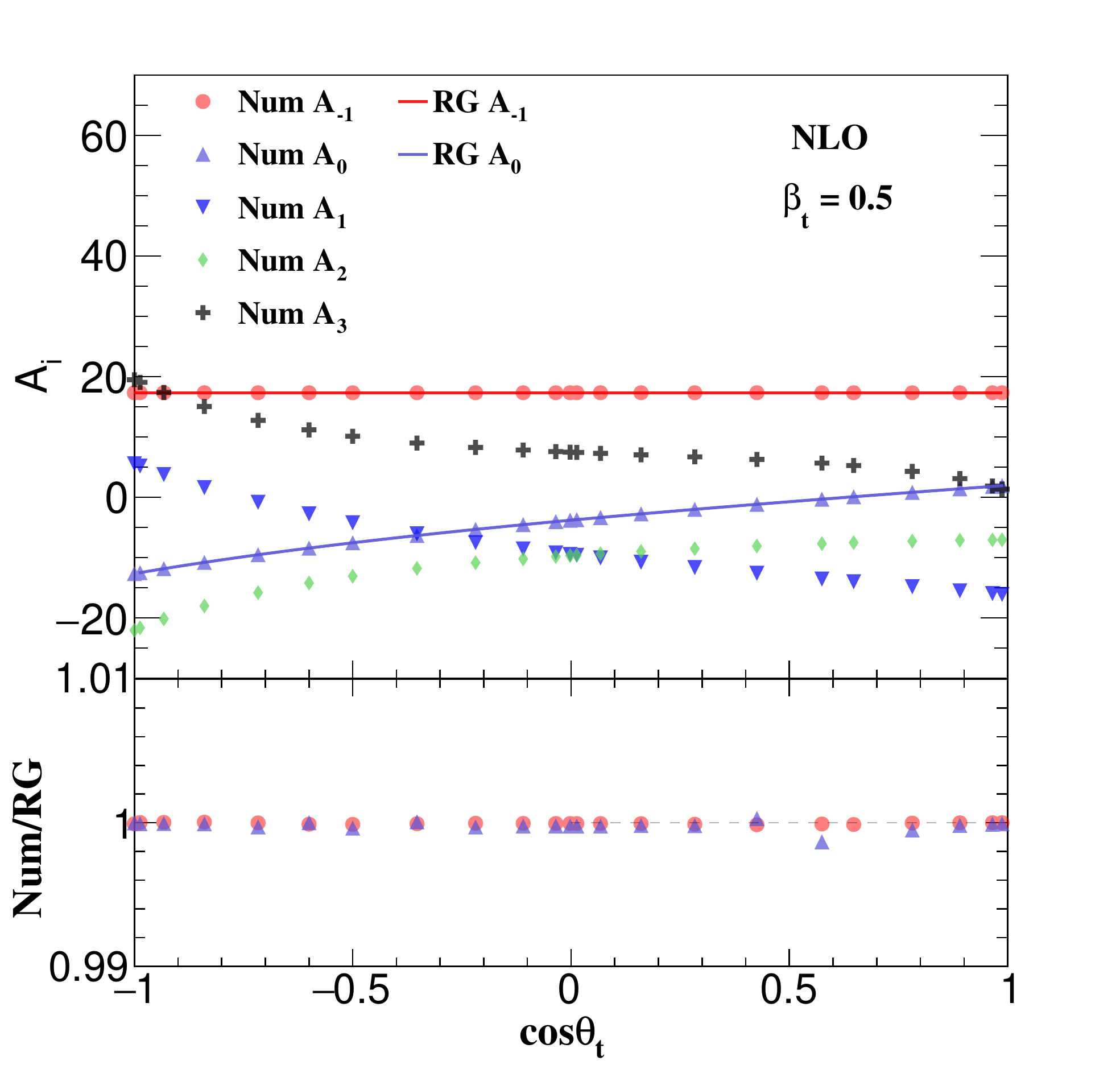}
    \includegraphics[scale=0.35]{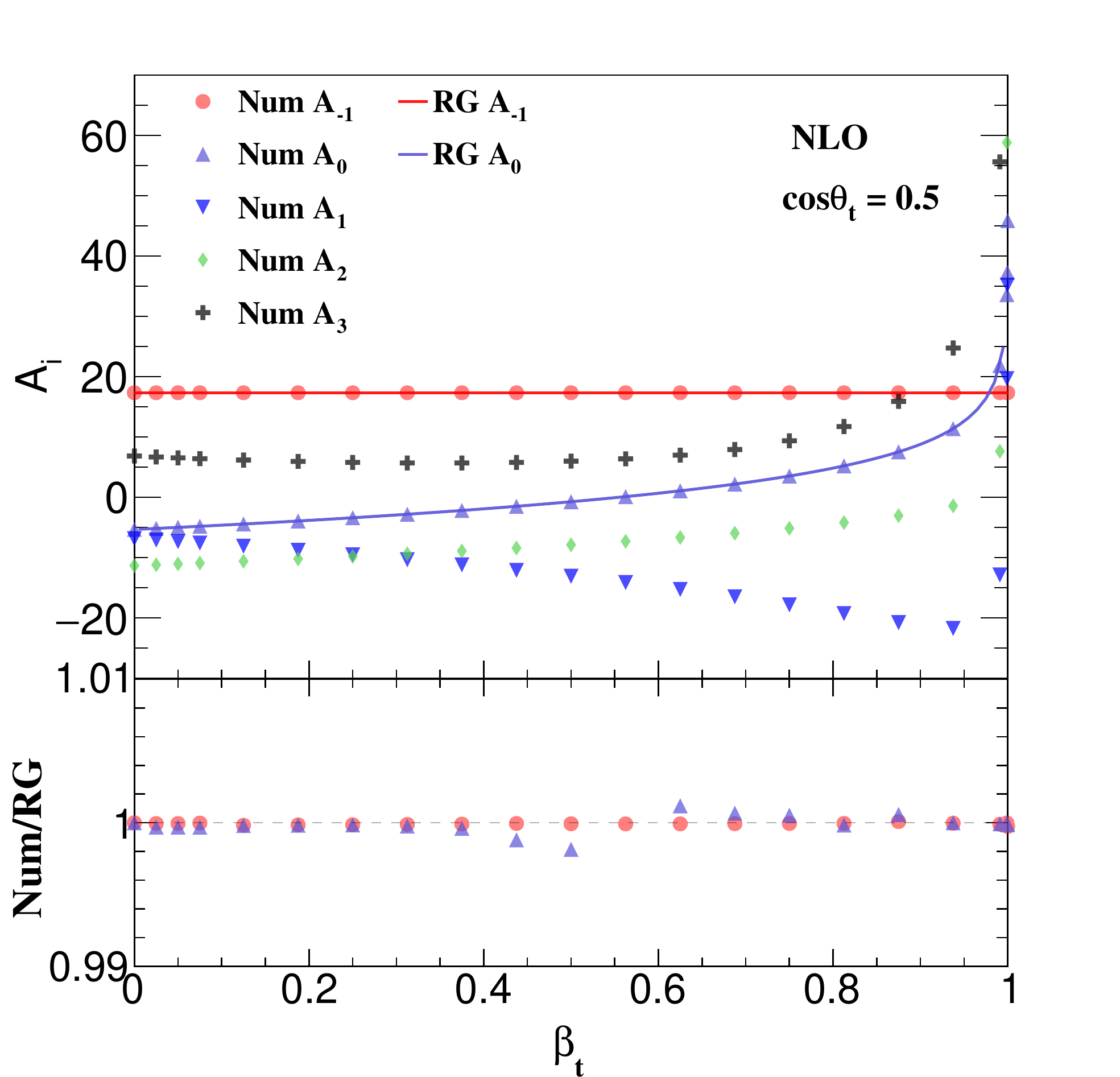}
    \caption{Numerical results for the NLO soft function and the comparison of $A_{-1}$ and $A_0$ with the RG predictions with fixed $\cos\theta_t$ (left) and $\beta_t$ (right).  }
    \label{fig:nlo}
\end{figure}

The LO soft function is trivial and has been given explicitly in eq.(\ref{eq:baresoft}). In this section, we present its NLO result. Expanding the soft Wilson lines in eq.(\ref{eq:soft}) in a series of the strong coupling, we obtain
the NLO soft function
\begin{align}
   S^{(1)}(\tau) =\frac{ 2 e^{\gamma_E \epsilon}\mu^{2\epsilon} }{\pi^{1-\epsilon}}
 \int d^dq \delta(q^2) J_a^{\mu (0)\dagger}d_{\mu\nu}(q) J_a^{\nu (0)}(q)F(n,\bar{n},q)~,
\end{align}
where $e^{\gamma_E \epsilon}$ is inserted because we use $\overline{\rm MS}$ renormalization scheme. The factor $J_a^{\mu (0)}(q)$ is the LO one-gluon soft current, or the eikonal current,
\begin{align}
    J_a^{\mu (0)}(q) = \sum_{i=1}^{3} \mathbf{T}_i^{a} \frac{p_i^{\mu}}{p_i \cdot q}
\end{align}
with $a$ the color index.

After performing the phase space integration, we obtain the NLO bare soft function
\begin{align}\label{eq:snlo}
    s^{(1)} = 
    \frac{A_{-1}}{\epsilon} + A_{0} + A_{1} \epsilon + A_{2} \epsilon^2 + A_{3} \epsilon^3  + \mathcal{O}(\epsilon^4) ,
\end{align}
where $A_i$ is a function of $\beta_t$ and $\cos\theta_t$.  Figure 1 shows the numerical results for the NLO soft function and the comparison of the divergent coefficients between the numerical calculations and RG predictions with fixed $\cos\theta_t$ or $\beta_t$.  
The deviations are not larger than 0.2\% except for the case of $|A_i|\to 0$~. 
The points at $\beta_t = 0$ just reproduce our previous results in ref.~\cite{Li:2016tvb}, as expected.
It can also be seen that when $\beta_t \to 1$, i.e., the top quark is highly boosted,  
the coefficients $A_i,i=0,1,2,3,$ become divergent.
This is due to the logarithmic structures such as $\ln^n (1-\beta_t)$ in the limit of $\beta_t \to 1$. 
In principle, this kind of logarithms can be predicted from effective field theory for boosted top productions.
Because the top quark mass is small compared with its energy in the limit, the scale hierarchy of the process is $\tau\ll m_t \ll \sqrt{\hat{s}}$,
and thus a different factorization formula should be derived.
We leave the detailed discussion to a future work.
Notice that in eq.~(\ref{eq:snlo}) and fig.~\ref{fig:nlo} we also show $A_2$ and $A_3$ which do not contribute to the NLO result.
However, they will contribute to the renormalized NNLO soft function.

\subsection{NNLO soft function} 

The NNLO contribution consists of two parts, i.e.,
\begin{align}
     s^{(2)} = s_{\rm VR}^{(2)}+s_{\rm DR}^{(2)}.
\end{align}
The first part is the virtual-real correction, i.e., the one-loop virtual corrections to LO soft gluon current $J^{\mu(1)}_a(q)$;
the second part is the double-real correction, i.e., the corrections with a double-gluon soft current $J^{\mu\nu(0)}_{ab}(q_1,q_2)$ or a massless quark-pair emission.
For the virtual-real contribution  we use  the soft limit of one-loop QCD amplitudes which has been studied
in refs.~\cite{Bern:1998sc,Bern:1999ry,Catani:2000pi} and ref.~\cite{Bierenbaum:2011gg} for massless and  massive external particles.  As for the double-real contribution we make use of the results in refs.~\cite{Catani:1999ss,Czakon:2011ve} where
the infrared behaviour of tree-level QCD amplitudes at NNLO has been analyzed. The details of the virtual-real and double-real matrix element  can be found in our previous paper~\cite{Li:2016tvb}. 

\begin{table}
\begin{tabular}{c|c|c|c|c|c|c|c|c }
\hline\hline
  & \multicolumn{2}{|c|}{$B_{-3}$} & \multicolumn{2}{|c|}{$B_{-2}$} &
  \multicolumn{2}{|c|}{$B_{-1}$} &
  \multicolumn{2}{|c}{$B_{0}$} 
  \\
  \cline{2-9}
  &  Num & RG &  Num & RG &  Num & RG &  Num & RG 
  \\ \hline
  $C_A^2$ & -8.0000 & -8  & 2.4972  & 2.4968 & 84.3749 & 84.3784  & 147.222 &147.233 
    \\ \hline
  $C_F^2$ &-8.0004 & -8 & 19.6943 &19.6916  & 26.6838 & 26.6908 &64.0611  &63.9972 
    \\ \hline
  $C_A C_F$ & -16.0000 & -16  &22.1903 & 22.1885  &107.408 & 107.386  &194.270  &194.217
    \\ \hline
  $C_A n_f $ & 0 & 0 &-1.3332 & -1.3333  &-3.0273 &-3.0283  &3.2803 &3.2779
      \\ \hline
  $C_F n_f $ & 0 & 0  & -1.3335 & -1.3333   &1.0599  &1.0597  & -1.0928 &-1.0949
  \\ \hline
  Max.devi.  & \multicolumn{2}{|c|}{$5\times 10^{-5}$} & \multicolumn{2}{|c|}{$1.5\times 10^{-4}$} &
  \multicolumn{2}{|c|}{$3.5\times 10^{-4}$} &
  \multicolumn{2}{|c}{$1.3\times 10^{-3}$} 
  \\
  \hline \hline 
\end{tabular}
\caption{Comparison between the numerical calculations and the RG predictions of the divergent terms in different color factors with $\beta_t=0.3$ and $\cos\theta_t = 0.5 $.
In the last line, we show the maximum deviation of the numerical calculations with respect to the RG predictions.}
\label{tab:nnloo}
\end{table}

\begin{figure}
    \centering
    \includegraphics[scale=0.35]{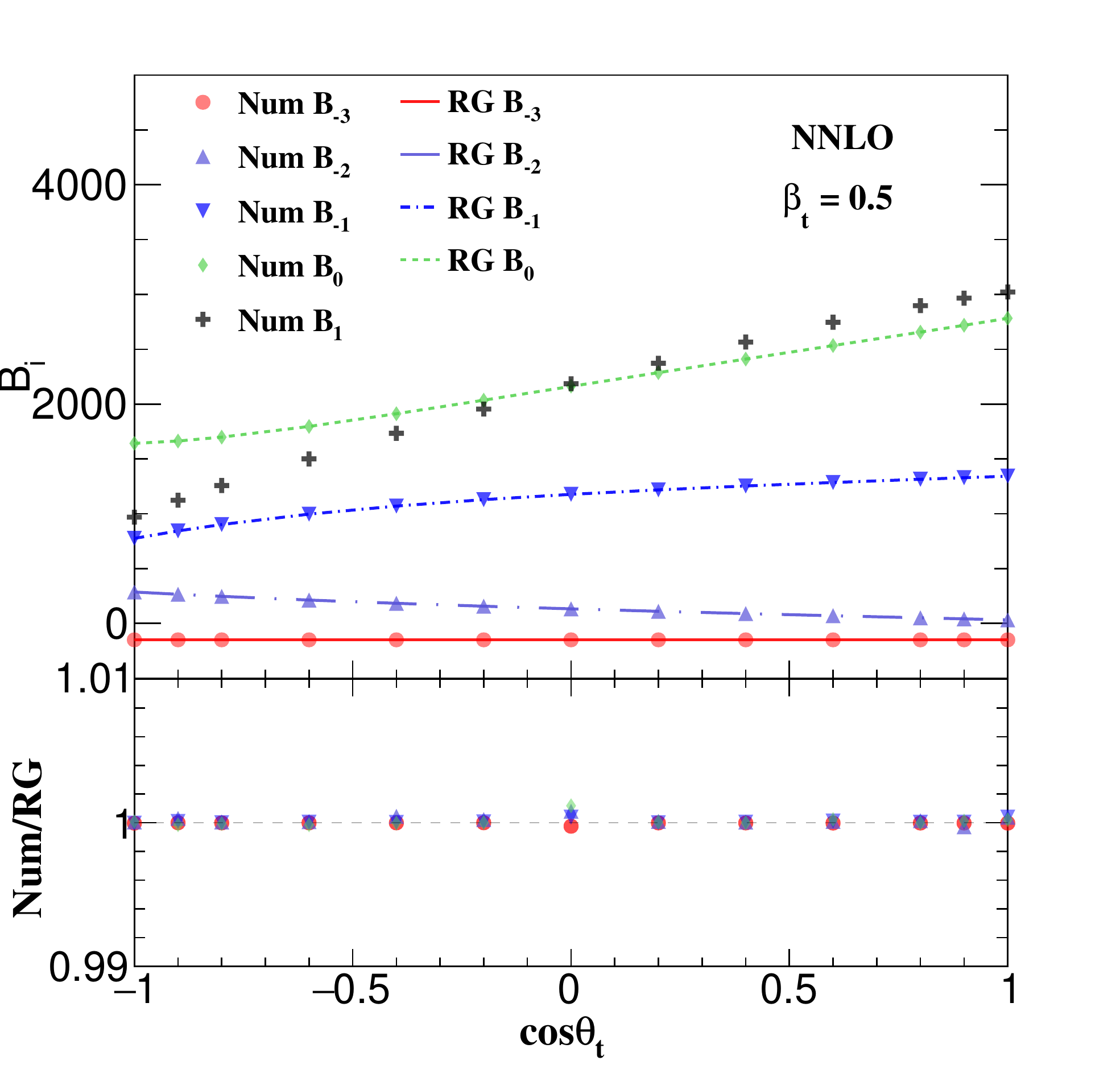}
    \includegraphics[scale=0.35]{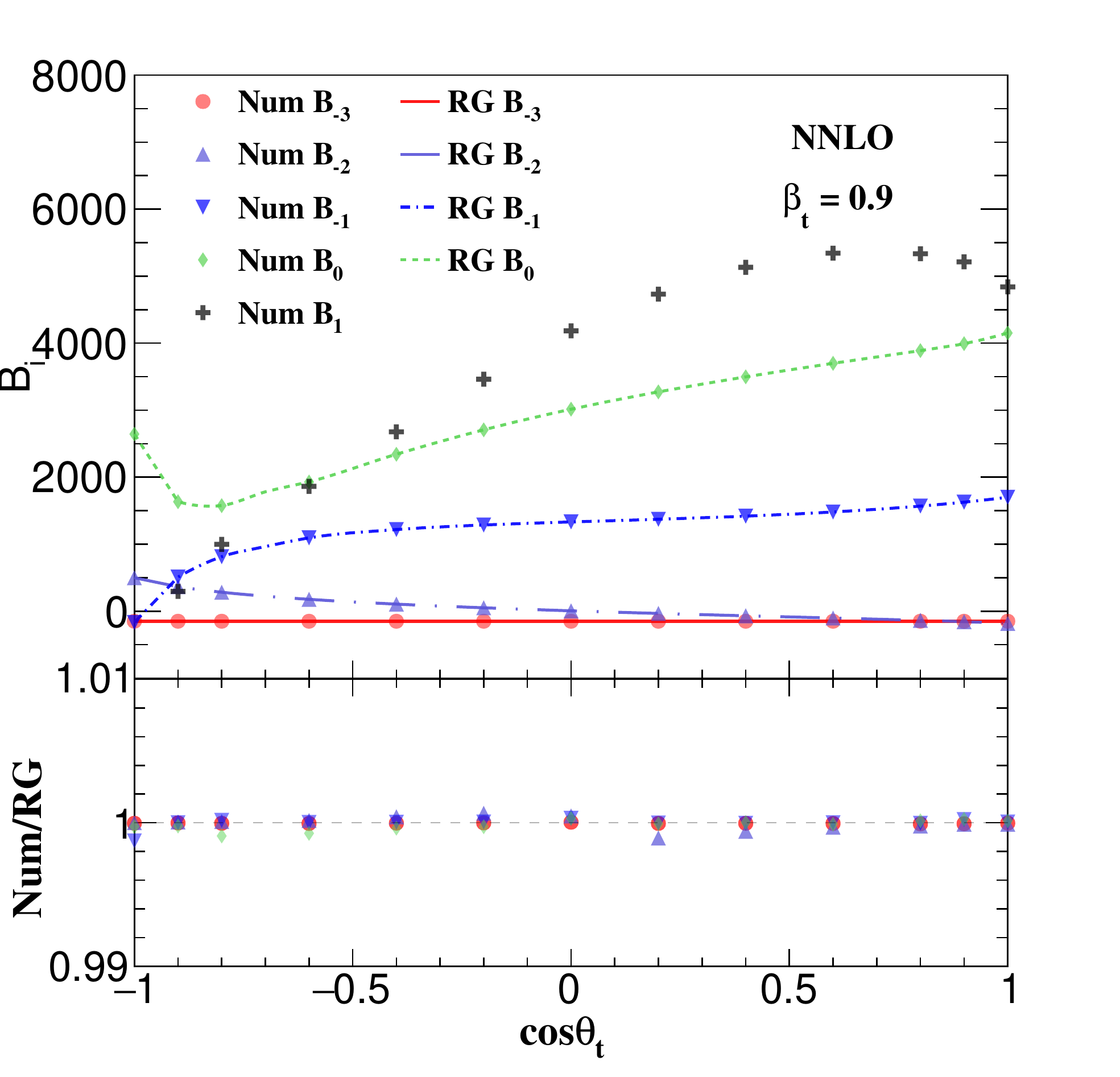}  
    \caption{Numerical results for NNLO bare soft function and the comparison to the RG predictions with $\beta_t =$ 0.5 (left) and 0.9 (right). The color factors are $C_A=3$ and $C_F=4/3$ and the number of flavors is  $n_f=5$.  }
    \label{fig:nnlo_betat}
\end{figure}

\begin{figure}
    \centering
    \includegraphics[scale=0.35]{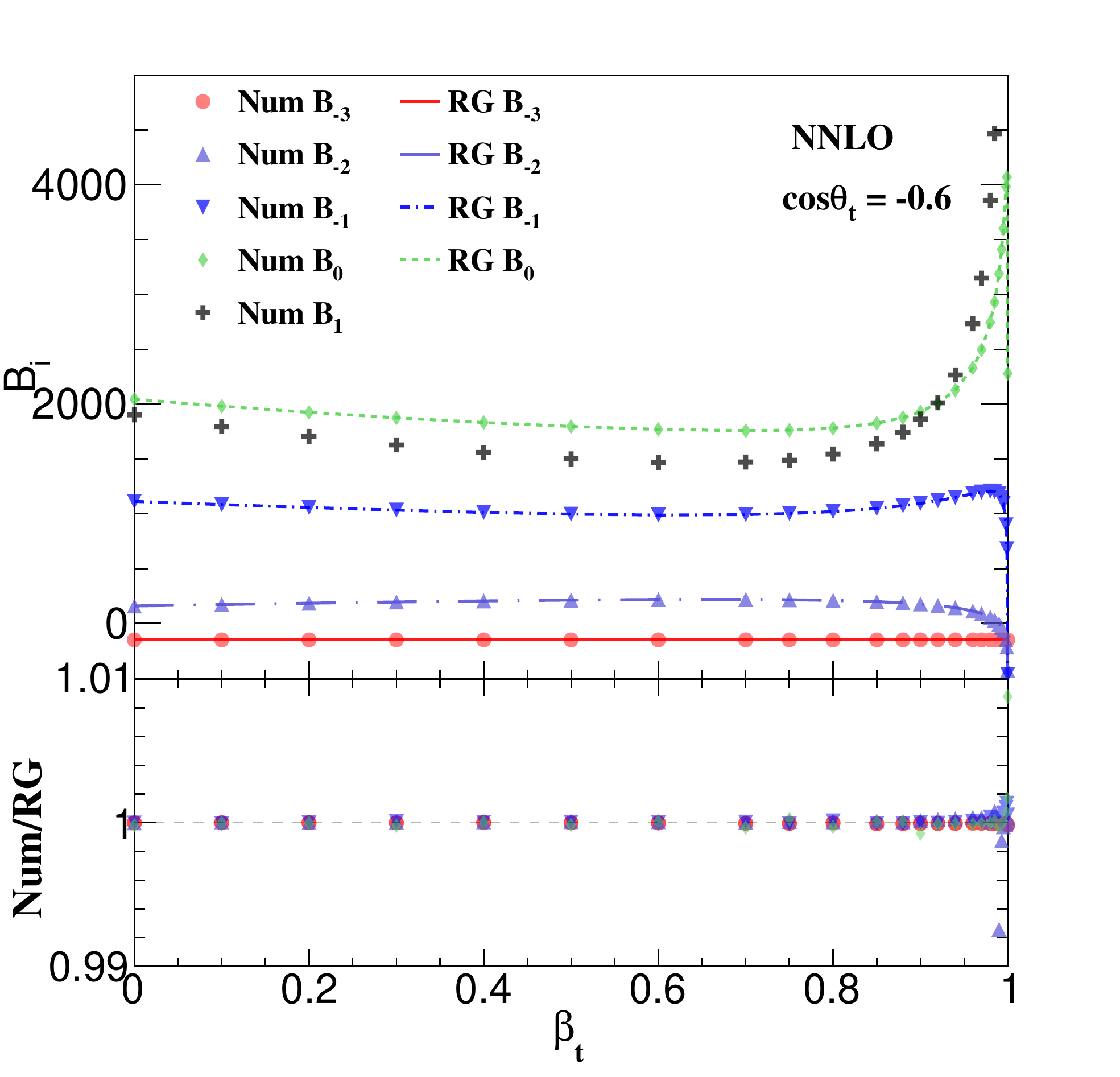}
    \includegraphics[scale=0.35]{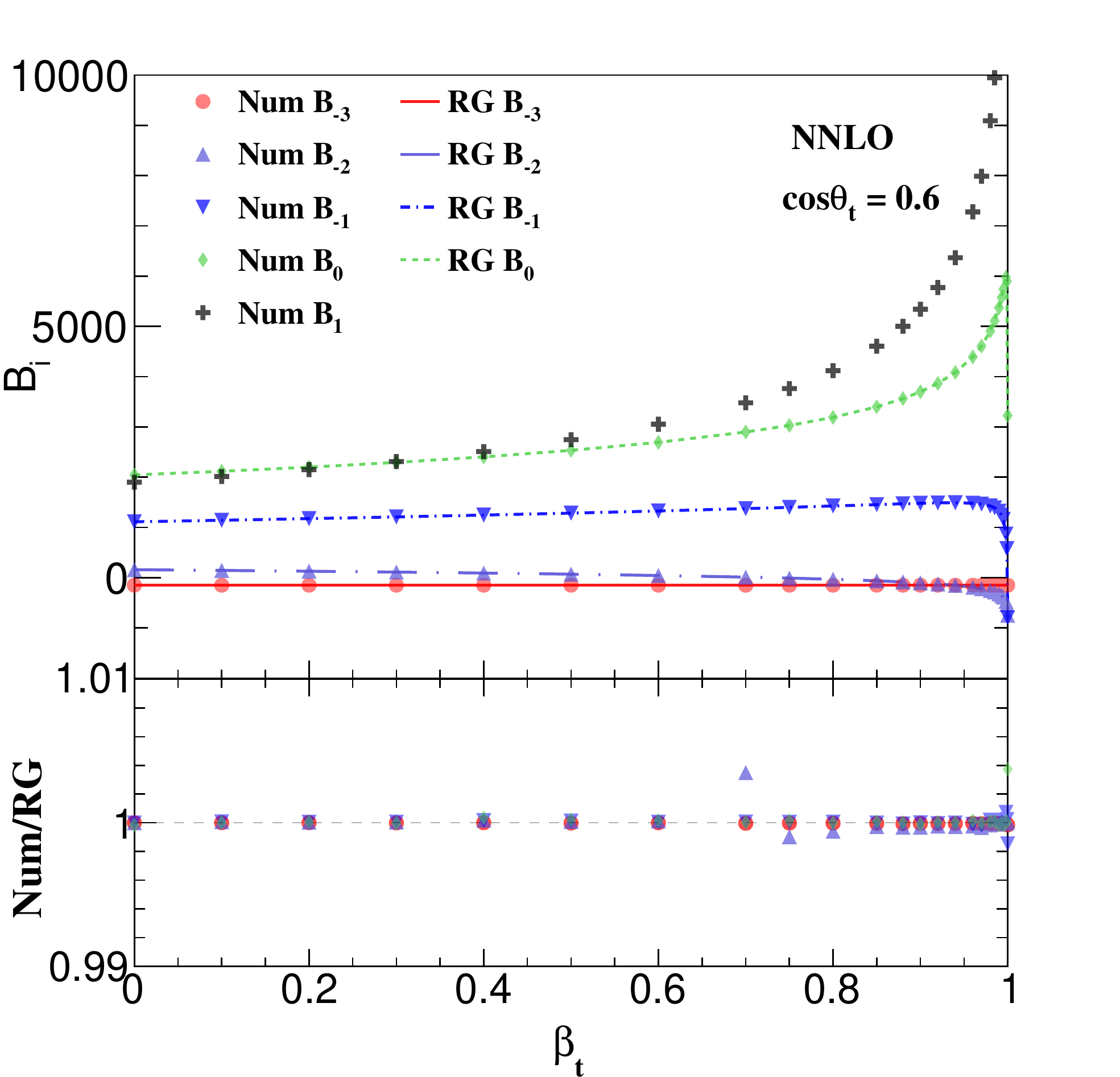}  
    \caption{Numerical results for NNLO bare soft function and the comparison to the RG predictions with $\cos\theta_t =$ -0.6 (left) and 0.6 (right). The color factors are $C_A=3$ and $C_F=4/3$ and the number of flavors is  $n_f=5$. }
    \label{fig:nnlo_ct}
\end{figure}

With the techniques discussed in section \ref{sec:tech}, the double-real part is calculated numerically after sector decomposition. The bare soft function  at NNLO, defined in eq.~(\ref{eq:baresoft}), can be written as 
\begin{align}
    s^{(2)} = \frac{B_{-3}}{\epsilon^3}
    +\frac{B_{-2}}{\epsilon^2}
    +\frac{B_{-1}}{\epsilon^1}
    +B_0 + B_1 \epsilon  + \mathcal{O}(\epsilon^2)~.
\end{align}
Using eq.~(\ref{eq:s_rg}) and the anomalous dimensions in eq.~(\ref{eq:gammas})  the divergent terms in the bare NNLO soft function can be predicted, which is an important cross check of our calculations. 

Table~\ref{tab:nnloo} shows the comparison of the divergent terms in different color structures 
with fixed $\beta_t=0.3$ and $\cos\theta_t=0.5$. We see that the maximum deviation is less than $0.2\%$.
Figures~\ref{fig:nnlo_betat} and \ref{fig:nnlo_ct} show the numerical calculations and the RG  predictions 
with  $\cos\theta_t$ in the range of $(-1,1)$ but fixed $\beta_t$ and 
with $\beta_t$ in the range of $(0,1)$ but fixed $\cos \theta_t$, respectively. 
We find that the numerical results are consistent with the RG  predictions.  For most of the cases the deviations are less than $0.2\%$, while the deviations can be about $1\%$ only when the absolute values of the coefficient $B_i$ are close to zero.  
We have checked that the points at $\beta_t = 0$ reproduce our previous results in ref.~\cite{Li:2016tvb}.
Similar to the NLO results, in the highly boosted region, the NNLO coefficients contains logarithmic structures such as $\ln^n(1-\beta_t)$. 
They are divergent when $\beta_t \to 1$.
This fact explains the behaviour of the distributions of the points near the end point of $\beta_t$ in fig.\ref{fig:nnlo_ct}.

\section{Conclusions}
\label{sec:conc}
The $N$-jettiness subtraction method is one of the efficient methods to perform differential calculations of the NNLO cross sections. 
In this paper, we present the calculation of NNLO soft function for 
 one massive colored particle production which is one of the  indispensable ingredients in $N$-jettiness subtraction method.
Our calculation makes use of the one-loop soft current and infrared limit of the QCD matrix elements from refs.~\cite{Catani:2000pi,Bierenbaum:2011gg,Catani:1999ss,Czakon:2011ve} to construct the integrand. The phase space integrations are performed with the sector decomposition method and the techniques are discussed in details.  The divergent terms of NLO and NNLO soft functions in our calculations are in very good agreement with those from the RG predictions. Though our result is general for a single massive colored particle production,
we focus on $tW$ production in the discussion because it is one of the most important processes in the SM.
Once the two-loop hard function is obtained,  we can perform the NNLO calculation for the differential cross section of $tW$ production at hadron colliders.  
Our method can also be applied to the calculation of the $N$-jettiness soft function for top quark pair production,
which provides another way to study the NNLO differential cross section for this process. 
We leave this application in future study. 

\section*{Acknowledgements}
HTL would like to acknowledge the TU Munich for its hospitality during the completion of this work. 
We thank Xiaohui Liu for useful discussion.
The work of HTL was supported by Department of Energy Early Career Program.
The work of JW was supported  by the BMBF project No. 05H15WOCAA.
\bibliography{softpaper}
\bibliographystyle{JHEP}

\end{document}